\documentclass[a4paper,10pt]{article}

\usepackage{epsfig}
\usepackage{graphicx}
\usepackage{indentfirst}
\usepackage{amsmath}
\usepackage{amsfonts}
\usepackage{amssymb}

\begin{document}
\begin{center}
{\Large\bf Interactions of some fluids with dark energy in $f(T)$ theory}\\

\medskip

S. B. Nassur$^{(a)}$\footnote{e-mail:nassurmaeva@gmail.com}, M. J. S. Houndjo$^{(a,b)}$\footnote{e-mail: sthoundjo@yahoo.fr}, 
I. G. Salako$^{(a,c)}$\footnote{e-mail: inessalako@gmail.com} and J. Tossa$^{(a)}$\footnote{e-mail: joel.tossa@imsp-uac.org}

$^a$ \,{\it Institut de Math\'{e}matiques et de Sciences Physiques (IMSP)}\\
 {\it 01 BP 613,  Porto-Novo, B\'{e}nin}\\
$^{b}$\,{\it Facult\'e des Sciences et Techniques de Natitingou - Universit\'e de Natitingou - B\'enin} \\
$^{c}$\,{ D\'epartement de Physique, Universit\'e de K\'etou, K\'etou, B\'enin}

\date{}

\end{center}
\begin{abstract}
We investigate the interaction of the dark energy with some fluids filling the universe in the framework of $f(T)$ theory, where $T$ denotes the torsion scalar, searching for the associated gravitational actions. Dark energy is assumed to be of gravitational origin. The interaction of dark energy and baryonic matter is considered resulting in a decay of the energy density of the ordinary matter, where universe appears as driven by cosmological constant. Furthermore we consider the interaction of dark energy with Van Der Waals fluid and in this regard the universe is seemed plunging into a phantom phase. Finally, the interaction of the dark energy with Chaplygin gas is studied, leading to a universe dominated by the cosmological constant.
\end{abstract}

\section{Introduction}
It is well known that our universe is now experiencing an accelerated expansion. Several cosmological observational data defend this phenomenon as supernovae type Ia, cosmic microwave background radiation, large scale structure, baryon acoustic oscillations and weak lensing \cite{1001,1de1207.1646,2de1207.1646,Ries}. Relying astronomical observational data, the universe is composed about 73\% of dark energy and 27\% of dark matter plus baryonic matter\cite{Brevik}.\par
Dark energy and dark matter are thought to be respectively responsible for the current acceleration of the universe and of the dynamics of spiral galaxies. Their presence in the universe is one of the puzzles of theoretical physics. Thus, in this framework, there are a number of approaches dedicated to realize the effects of these mysterious components of the universe. Among these approaches, there may be mentioned the QCD dark energy what has been proposed to interpret the dark energy without any new parameter nor new degree of freedom. In this model, the squared sound speed of the dark energy is negative, indicating an instability of the model against perturbation theory
\cite{In.Mod.Phys.D}. However the most famous approaches wanting demystified the effects of the black components of our universe, are those that to ``reconsider'' the cosmological constant $\Lambda$ in Einstein's equations of general relativity (GR). The energy associated with the cosmological constant represents the energy of the quantum vacuum\cite{Padmanabhan} and it is associated to negative pressure which is the parameter of equation of state $\omega_{\Lambda}=-1$ . Another popular approach is to modify the  RG by replacing its geometric Lagrangian density $R$ by algebraic function $f(R)$ where $R$  is the variable. This approach leads to the $f(R)$ gravity(see \cite{Nojiri}).\par 
Gravity is one of the four natural fundamental interactions. It is inscribed in space-time by either RG or by teleparallel theory (TERG). General relativity can be defined as a geometric theory of gravitation, constructed from the Levi-Civita connection while the TERG can be defined as a gauge theory of gravity, from the connection Weizenbock \cite{Arcos}. Despite this basic difference, these two theories can be found to be equivalent \ cite {A.sousa, Shirafuji}, where the designation of teleparallel theory by teleparallel equivalent of general relativity (TERG). Thus the TEGR is also seen modified by including an algebraic function $f$ of the torsion scalar $T$, where $T$ is the geometric Lagrangian density of the TERG. This gives rise to the famous $f(T)$ gravity\cite{Rafael}.\par
Interesting results are obtained with $f(R)$ gravity\cite{Kauzar}, however, mathematical difficulties are more acute in this theory because its equations are four order. Unlikely, the $f(T)$ gravity has the advantage of producing two-order equations and it presents encouraging results (for some recent reviw see\cite{Nassur}).\par 
Dark energy has spilled much ink in the literature, seen as cosmological constant \cite {Sahni,Padmanabhan} or look in terms of fluid  pressure which is related to energy density in particular form. In this vision, many models are reconstructed in the literature and they can be seen as unifying dark energy and dark matter. In this paper, we consider two of these special fluids: van der Waals fluid and Chaplygin gas.\par
The Van der Waals equation of state was introduced to overcome the inconsistency which may arise in a thermodynamics analysis of a system in which two phases exist simultaneously. This coexistence of phases occurred including the equivalence between the radiation dominated phase and the dust dominated phase. Approaches were used by S. Capozziello et al.\cite{Capozziello,Capozziello1} to constrain the van der Waals fluid to observational data. In these studies, it was shown that the van der Waals fluid can be seen as unifying dark energy and dark matter. In addition, G. M. Kremer analyzes the van der Waals fluid \cite{Kremer} by considering it at first as the only content of the universe and in second time as an additional component of the universe in more dark energy.\par 
Analysis of Chaplygin gas made by N. Bilic et al. \cite{Bilic} led to the conclusion on the effective unification of dark energy and dark matter with this gas, however, this unification is closely related to the geometry of space (its inhomogeneous character). A study of the Chaplygin gas through the setting of the state-finder parameters was discussed by V. Gorini et al.\cite{Gorini}. Furthermore a global stability of Chaplygin gas was shown by V. Gorini et al.\cite{phys.rev.d}. On the other hand, the structural stability of Chaplygin gas was analyzed by M. Szydlowski and W. Czaja\cite{Melek}.\par 
It is customary that by a cosmological analysis in $ f(T)$ gravity, one makes a choice either on the consideration of a Lagrangian density $f(T)$ and one sees the behavior of cosmological parameters by determining the scale factor $a(t)$, or by providing the scale factor and one determines the action $f(T)$ by seeing the behavior of the cosmological parameters. Our work would follow the second approach with {\it particularity} i.e the aim of this work is to reconstruct the action $f(T)$ by interacting whenever two fluids that one considers as the only content of the universe. Several studies taking into account the fluid interactions supposed form the contents of the world have been addressed on different theories (see \cite{Nojiri1,Nojiri2,In.Mod.Phys.D,Brevik,Rudra}). {\it The particularity}
that we are referring here is the fact that our approach although follows the second point, the reconstructed actions does not depend on the chosen scale factor, as it is often the case. The chosen scale factor is just a tool for the cosmological analysis.\par 
We will begin this work by the generality of $f(T)$ gravity in Section \ref{sec2}. Then the approach of interactions will start by the interaction of dark energy and dark matter in vacuum,in section \ref{sec3}. The section \ref{sec4} will be devoted to the interaction of dark energy and baryonic matter. The Van der Waals fluid and dark energy will be studied in Section \ref {sec5}. The section \ref{sec6} is reserved to the interaction of dark energy and Chaplygin gas. And finally the conclusion will be given in section \ref{sec7}.

\section{Generality}\label{sec2}
The modified theory of gravity based on the torsion scalar is the one for which the geometric part of the action is an algebraic function depending on the torsion. In the same way as in the Teleparallel gravity, the geometric elements are described using orthonormal tetrads components defined in the tangent space at each point of the manifold. In general the line element can be written as
\begin{eqnarray}
ds^2=g_{\mu\nu}dx^\mu dx^\nu=\eta_{ij}\theta^i\theta^j\,,
\end{eqnarray}
where we define the following elements
\begin{eqnarray}
dx^\mu=e_{i}^{\;\;\mu}\theta^{i}\,\quad \theta^{i}=e^{i}_{\;\;\mu}dx^{\mu}.
\end{eqnarray}
Note that $\eta_{ij}=diag(1,-1,-1,-1)$ is the metric related to the  Minkowskian spacetime and the $\{e^{i}_{\;\mu}\}$ are the components  of the tetrad which satisfy the following identities
\begin{eqnarray}
e^{\;\;\mu}_{i}e^{i}_{\;\;\nu}=\delta^{\mu}_{\nu},\quad e^{\;\;i}_{\mu}e^{\mu}_{\;\;j}=\delta^{i}_{j}.
\end{eqnarray}
The connection in use in this theory is the one of  Weizenbock's,  defined by
\begin{eqnarray}
\Gamma^{\lambda}_{\mu\nu}=e^{\;\;\lambda}_{i}\partial_{\mu}e^{i}_{\;\;\nu}=-e^{i}_{\;\;\mu}\partial_\nu e_{i}^{\;\;\lambda}.
\end{eqnarray}
Once the previous connection is assumed, one can then express the main geometric objects; the torsion tensor's components as\begin{eqnarray}
T^{\lambda}_{\;\;\;\mu\nu}= \Gamma^{\lambda}_{\mu\nu}-\Gamma^{\lambda}_{\nu\mu},
\end{eqnarray}
which is used in the definition of the contorsion tensor as
\begin{eqnarray}
K^{\mu\nu}_{\;\;\;\;\lambda}=-\frac{1}{2}\left(T^{\mu\nu}_{\;\;\;\lambda}-T^{\nu\mu}_{\;\;\;\;\lambda}+T^{\;\;\;\nu\mu}_{\lambda}\right)\,\,.
\end{eqnarray}
The above objects (torsion and contorsion) are used to define a new tensor $S_{\lambda}^{\;\;\mu\nu}$ as
\begin{eqnarray}
S_{\lambda}^{\;\;\mu\nu}=\frac{1}{2}\left(K^{\mu\nu}_{\;\;\;\;\lambda}+\delta^{\mu}_{\lambda}T^{\alpha\nu}_{\;\;\;\;\alpha}-\delta^{\nu}_{\lambda}T^{\alpha\mu}_{\;\;\;\;\alpha}\right)\,\,.
\end{eqnarray}
The torsion scalar is defined from the previous tensor and the torsion tensor as
\begin{eqnarray}
T=T^{\lambda}_{\;\;\;\mu\nu}S^{\;\;\;\mu\nu}_{\lambda}
\end{eqnarray}
Let us write the action for the modified $f(T)$ theory with matter as follows
\begin{equation}
\label{1}
 S=\int d^{4}x e\left[\frac{\mathcal{L}_{G}}{2\kappa^{2}}+\mathcal{L}_{(matter)}\right],
\end{equation}
where $\kappa^{2}=8\pi G$, $e\equiv\det[e^{i}\,_{\mu}]=\sqrt{-g}$ denotes the determinant of the tetrad and $g$ the determinant of the space-time metric.\par
We consider the geometric Lagrangian density as
\begin{eqnarray}
 \mathcal{L}_{G}=T+f(T).
\end{eqnarray}
The equations of motion in the flat FRW universe are given by
\cite{Bamba,Karami}
\begin{eqnarray}
 6H^{2}+12H^{2}f_{T}(T)+f(T)=2\kappa^{2}\rho_{i},\\
 2(2\dot{H}+3H^{2})+f(T)+4(\dot{H}+3H^{2})f_{T}(T)-48H^{2}\dot{H}f_{TT}(T)=2\kappa^{2}p_{i},
\end{eqnarray}
where $\rho_{i}$ and $p_{i}$ denote the energy density and pressure associated with a type of perfect fluid. For $\kappa^{2} =8\pi G\equiv 1$, the above equations can be rewritten as
\begin{eqnarray}
\label{fr1}
 H^{2}=\frac{1}{3}(\rho_{i}+\rho_{T}),\\
 \label{fr2}
 \dot{H}=-\frac{1}{2}(p_{i}+\rho_{i}+p_{T}+\rho_{T}).
\end{eqnarray}
With \cite{Karami}
\begin{eqnarray}
\label{dem}
 \rho_{T}=-\frac{1}{2}(12H^{2}f_{T}+f),\\
 \label{pem}
 p_{T}=\frac{1}{2}\left(4\dot{H}(1+f_{T}-12H^{2}f_{TT})+12H^{2}f_{T}+f\right).
\end{eqnarray}
Here, the relations (\ref{fr1}) and (\ref{fr2}) are constraints equations. They will be used to determine the integration constants.
Furthermore, equation (fr1) may be described as follows
\begin{eqnarray}
 \Omega_{eff}=1,
\end{eqnarray}
where
\begin{eqnarray}
\label{fr3}
 \Omega_{eff}\equiv\Omega_{i}+\Omega_{T}\equiv\frac{\rho_{i}}{3H^{2}}+\frac{\rho_{T}}{3H^{2}}.
\end{eqnarray}
One poses
\begin{eqnarray}
 \rho_{eff}=\rho_{i}+\rho_{T},\\
 p_{eff}=p_{i}+p_{T},\\
\omega_{eff}=p_{eff}/\rho_{eff}.
\end{eqnarray}
For different graphs, we will use the scale factor introduced by S. Nojiri et al. as unifying a dominated ordinary matter phase and dominated dark energy phase \cite{Odintsov2,Odintsov1}
\begin{eqnarray}
 a=a_{0}t^{g_{1}}e^{g_{0}t},
\end{eqnarray}
with $a_{0}, g_{0}$ and $g_{1}$ being positive constants.\\
Now we will start our approach with the dark energy and dark matter.
\section{Interaction between dark energy and dark matter:``in vacuum''}\label{sec3}
One looks at the interaction between dark energy and dark matter, so in this section we will ignore on all other form of energy or matter. Thus, one assumes dark energy and dark matter densities and their respective pressures coming from the energy density and the pressure induced by the torsion. In other words, one sets
\begin{eqnarray}
\label{deT}
 \rho_{T}=\rho_{DE}+\rho_{DM},\\
 p_{T}=p_{DE}+p_{DM},
\end{eqnarray}
where $\rho_{DE},\rho_{DM}, p_{DE}$ and $p_{DM}$ are densities of dark energy and dark matter and their respective pressures. Considering that dark energy and dark matter are two perfect fluids that interact following the equations below \cite{In.Mod.Phys.D,Brevik}.
\begin{eqnarray}
\label{contDE}
 \dot{\rho}_{DE}+3H(p_{DE}+\rho_{DE})=-Q,\\
 \label{contDM}
 \dot{\rho}_{DM}+3H(p_{DM}+\rho_{DM})=Q,\\
 \label{fr21}
 \dot{H}=-\frac{\kappa^{2}}{2}(\rho_{DE}+p_{DE}+p_{DM}+\rho_{DM}).
\end{eqnarray}
These equations are more general in the sense that these two fluids may be considered as non-zero pressure. The derivation of equation (\ref{dem}) with the decomposition (\ref{deT}) gives
\begin{eqnarray}
\label{ddEDEDM}
 \dot{\rho}_{DE}+\dot{\rho}_{DM}=\frac{\dot{T}}{2}\left(f_{T}+2Tf_{TT}\right).
\end{eqnarray}
Furthermore, from the combination of relations (\ref{contDE}) and (\ref{contDM}), and making use of the equation (\ref{fr21}), one gets 
\begin{eqnarray}
\label{ad}
 \dot{\rho}_{DE}+\dot{\rho}_{DM}=-\frac{\dot{T}}{2}.
\end{eqnarray}
By combining the equations (\ref{ddEDEDM}) and (\ref{ad}), one obtains a differential equation on $f$ reporting about the interaction of dark energy and dark matter in the vacuum.
\begin{eqnarray}
\label{actDEDM}
 \frac{d^{2}f}{dT^{2}}+\frac{1}{2T}\frac{df}{dT}+\frac{1}{2T}=0.
\end{eqnarray}
A trivial solution can be given by
\begin{eqnarray}
 \label{TT}
 f(T)=-T+C
\end{eqnarray}
where $C$ an interact constant. Using equation (\ref{fr1}) it follows $K=0$. This leads to
\begin{eqnarray}
\label{densite}
 p_{eff}=-\rho_{eff}=-3H^{2},
\end{eqnarray}
whether
\begin{eqnarray}
\label{peff2}
 \omega_{eff}=-1,\\
 \Omega_{eff}=1.
\end{eqnarray}
The nontrivial solutions of equation (\ref{actDEDM}) are given by
\begin{eqnarray}
\label{actionDEDM}
 f(T)=-T-2C_{0}\sqrt{-T}+C_{1},
\end{eqnarray}
where $C_{0}$ and $C_{1}$  are integrations constants. From the equation (\ref{fr1}), one gets $C_{1}=0$ with $C_{1}$ arbitrary constant. 
Then the effective equation of state parameter $\omega_{eff}$ and the effective energy density parameter $\Omega_{eff}$ become
\begin{eqnarray}
\label{peff1}
\Omega_{eff}=1,\\
\omega_{eff}=-1.
\end{eqnarray}
It should be noted that the action (\ref{actionDEDM}) leads to the same expression for the effective energy density (\ref{densite}). Thus, as can be seen with the expansions (\ref{peff1}) and (\ref{peff2}), these actions are equivalent.
This is not surprising because they come from the same source (\ref{actDEDM}). Thus, the Lagrangian densities (\ref{TT}) and (\ref{actDEDM}) can be used to make a cosmological study which we want to take into account the interaction between dark energy
and dark matter in a universe dominated by these two components.
\begin{figure}
\begin{center}
\includegraphics[height=8cm,width=10cm]{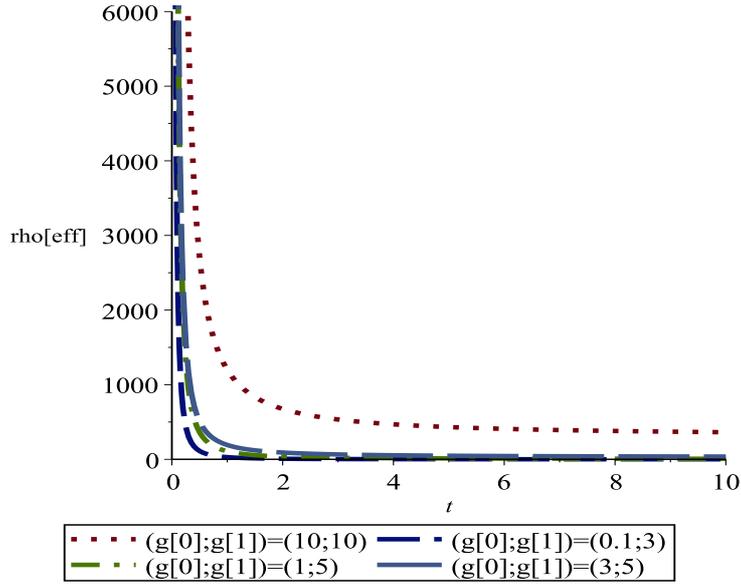}
\end{center}
\caption{\label{fig1} The graph presents the evolution of $\rho_{eff}$ versus $t$ for an universe devoid of common components where dark energy and dark matter interacting.}
\end{figure}
\section{Interaction between dark energy and ordinary matter}\label{sec4}
In this section as before we will look at two components as the only that dominate the universe, that is the case of dark energy and ordinary matter. Then we will assume from now that torsion induces a single component that is dark energy, i.e $\rho_{T}=\rho_{DE}$. Then the equations of interactions between dark energy and ordinary matter are given by \cite{Nojiri1,Nojiri2,Rudra}
\begin{eqnarray}
\label{contDE1}
 \dot{\rho}_{DE}+3H(p_{DE}+\rho_{DE})=-Q,\\
 \label{contM}
 \dot{\rho}_{ord}+3H(p_{ord}+\rho_{ord})=Q,\\
 \label{fr22}
 \dot{H}=-\frac{1}{2}(\rho_{DE}+p_{DE}+p_{ord}+\rho_{ord}).
\end{eqnarray}
Our approach also takes into account of an ordinary matter non zero pressure. The derivation of equation (\ref{dem}) gives
\begin{eqnarray}
\label{ddEDE}
 \dot{\rho}_{DE}=\frac{\dot{T}}{2}(2Tf_{TT}+f_{T}).
\end{eqnarray}
By combining relations (\ref{contDE1}) and (\ref{contM}) and using the equation (\ref{fr22}), one obtains
\begin{eqnarray}
\label{adEM}
 \dot{\rho}_{DE}+\dot{\rho}_{ord}=-\frac{\dot{T}}{2}.
\end{eqnarray}
By making use of the equations (\ref{ddEDE}) and (\ref{adEM}) one gets the following differential equation
\begin{eqnarray}
\label{actDEM}
 \frac{d^{2}f}{dT^{2}}+\frac{1}{2T}\frac{df}{dT}+\frac{1}{2T}=-\frac{1}{T}\frac{d\rho_{ord}}{dT}.
\end{eqnarray}
Equation (\ref{actDEM}) can be decomposed into two differential equations by the variables separation as follows
\begin{eqnarray}
 \label{EqactEDM}
 \frac{d^{2}f}{dT^{2}}+\frac{1}{2T}\frac{df}{dT}=E,\\
 \label{ddeo}
 \frac{1}{T}\frac{d\rho_{ord}}{dT}+\frac{1}{2T}=-E,
\end{eqnarray}
where $E$ is a non-zero constant. The solutions of these equations are written in the form
\begin{eqnarray}
 \label{act}
 f(T)=\frac{E}{3}T^{2}-2C_{2}(-T)^{1/2}+C_{3}
\end{eqnarray}
\begin{eqnarray}
 \label{eo}
 \rho_{ord}=-\frac{ET^{2}}{2}-\frac{T}{2}+C_{4},
\end{eqnarray}
with $C_{2},\, C_{3}$ and $C_{4}$  integration constants. By setting $\rho(t=t_{0})=\rho_{m0}$, $t_{0}$ being the present cosmic time cosmic  and $\rho_{m0}$ the current energy density of dust. One obtains $C_{4}=\rho_{m0}-3H^{2}_{0}(1+6EH^{2}_{0})$ where $H_{0}$ is the current  Hubble parameter. By using equation (\ref{fr1}), one gets $C_{3}=2C_{4}$ with $C_{2}$ an arbitrary constant.
The solution (\ref{act}) is the gravitational action in $f(T)$ theory with the account of the interaction between dark energy and ordinary matter for cosmological studies in a universe dominated by dark energy and ordinary matter. The effective equation of state parameter $\omega_{eff}$ and effective energy density $\Omega_{eff}$ parameter are given by
\begin{eqnarray}
 \omega_{eff}=-\frac{C_{3}+4\dot{H}(1+2ET)-ET^{2}}{T},\\
 \Omega_{eff}=1.
\end{eqnarray}
For $\rho_{m0}=3*1.5*10^{-67}eV^{2}$\cite{Odintsov3} and $H_{0}=74.2\pm3.6Km s^{-1} Mpc^{-1}$, one has
\begin{figure}[htbp]
\centering
\begin{tabular}{rl}
\includegraphics[height=7cm,width=7cm]{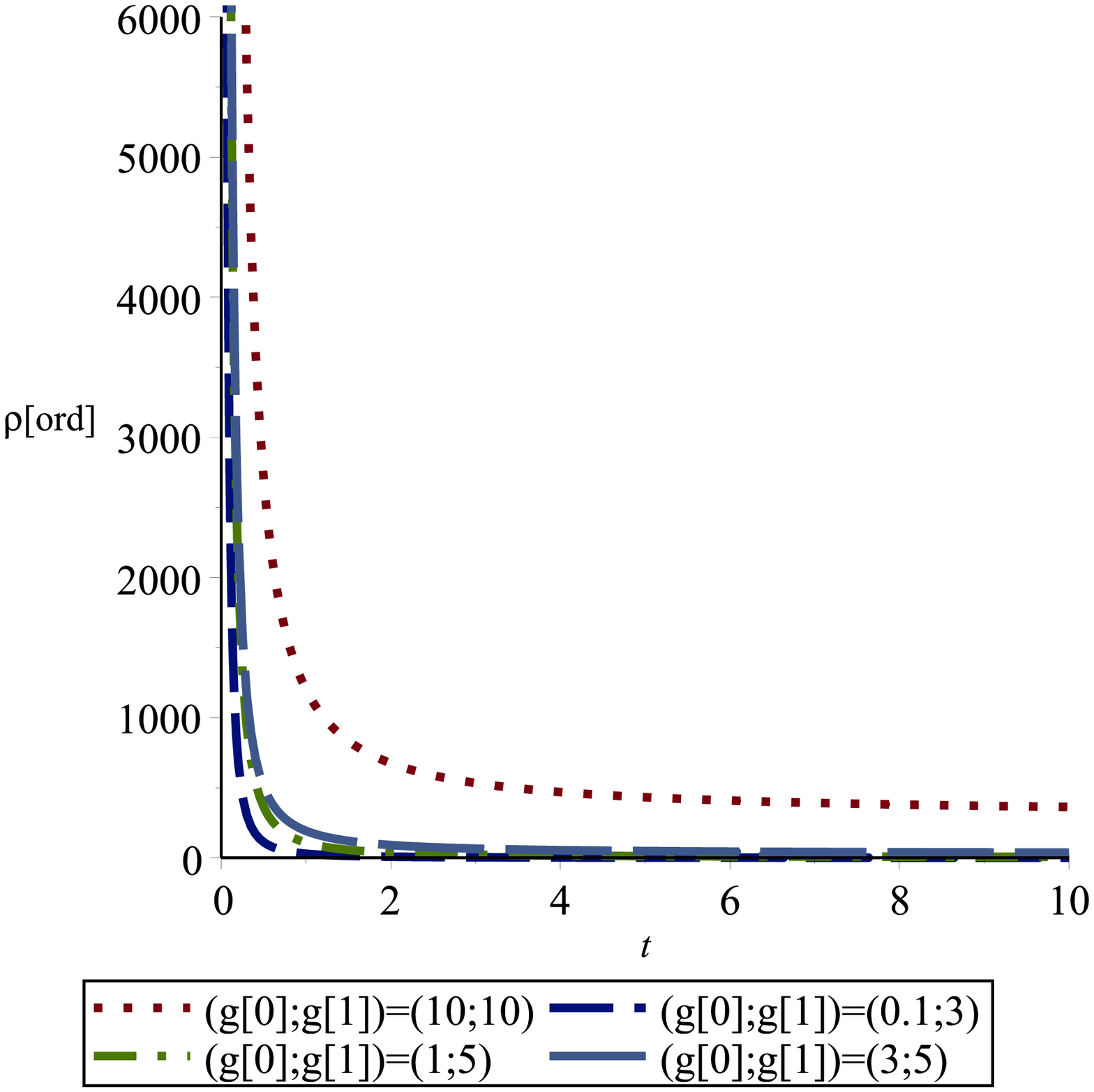} \label{e1}&
\includegraphics[height=7cm,width=7cm]{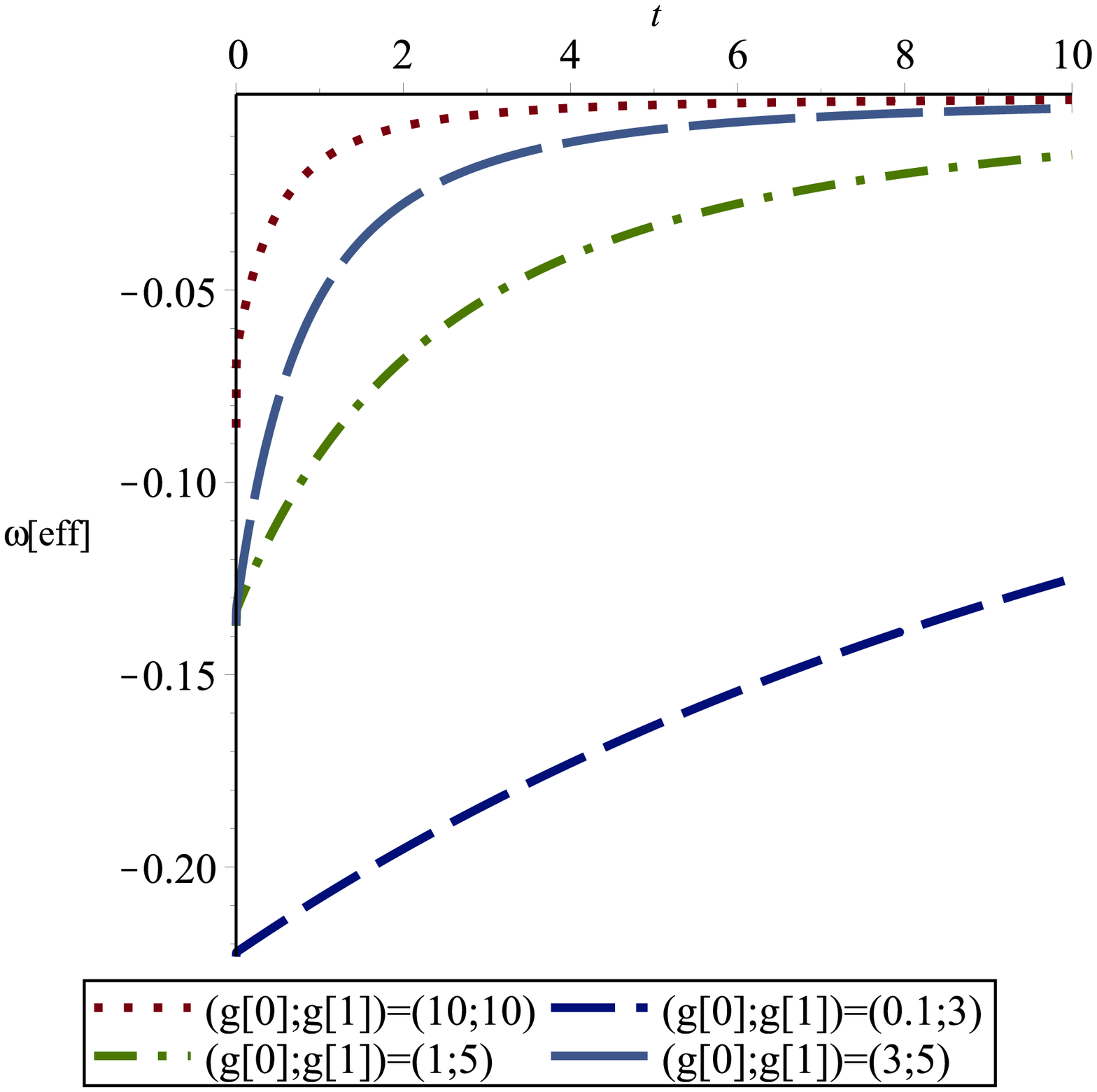} \label{e2}
\end{tabular}
\caption{The graphs present the evolution of $\rho_{ord}$ (left panel) and $\omega_{eff}$ (right panel) versus $t$ in an universe composed of dark energy and ordinary matter, interacting, for $E=10^{-10}$}
\label{fig2}
\end{figure}

\section{Interaction between dark energy and van der Waal fluid}\label{sec5}
Van der Waals fluid is characterized by its pressure $p_{w}$ which is given by\cite{Kremer}
\begin{eqnarray}
 p_{w}=\frac{8\omega_{w}\rho_{w}}{3-\rho_{w}}-3\rho^{2}_{w}.
\end{eqnarray}
The equations of interaction between dark energy and Van der Waals fluid can be given by
\begin{eqnarray}
 \dot{\rho}_{DE}+3H(p_{DE}+\rho_{DE})=-Q,\\
 \label{contW}
 \dot{\rho}_{w}+3H(p_{w}+\rho_{w})=Q,\\
 \label{dew}
 \dot{H}=-\frac{\kappa^{2}}{2}(\rho_{DE}+p_{DE}+p_{w}+\rho_{w}).
\end{eqnarray}
Using the same procedure as in section \ref{sec4}, one finds
\begin{eqnarray}
 \rho_{w}=-\frac{ET^{2}}{2}+C_{5},\\
 f(T)=\frac{ET^{2}}{3}-T+\frac{C_{6}\sqrt{-T}}{2}+C_{7}.
\end{eqnarray}
One considers $\dot{H}(t_{0})=0$ and $p_{DE}(t_{0})=-\rho_{DE}(t_{0})$, from equation (\ref{dew}) it comes that $p_{w}(t_{0})=-\rho_{w}(t_{0})$ whether $C_{5}=\frac{5}{3}-3EH_{0}^{2}+\sqrt{74-96\omega_{w}}$, for
$74-96\omega_{w}\geq0$. Using equation(\ref{fr1}), one obtains $C_{7}=2C_{5}$. So the effective EoS parameter $\omega_{eff}$ and the effective energy density parameter read 
\begin{eqnarray}
 \omega_{eff}&=&-1-\frac{1}{T}\left(8E\dot{H}T-ET^{2}+C_{7}\right.\\
 &&\left.-\frac{(ET^{2}+2C_{5})\left(32\omega_{w}+18ET^{2}-36C{5}+3E^{2}T^{4}-12ET^{2}C_{5}+4C_{5}^{2}\right)}{ET^{2}-2C_{5}+6}\right),\nonumber\\
 \Omega_{eff}&=&1.
\end{eqnarray}
\begin{figure}[htbp]
\centering
\begin{tabular}{rl}
\includegraphics[height=7cm,width=7cm]{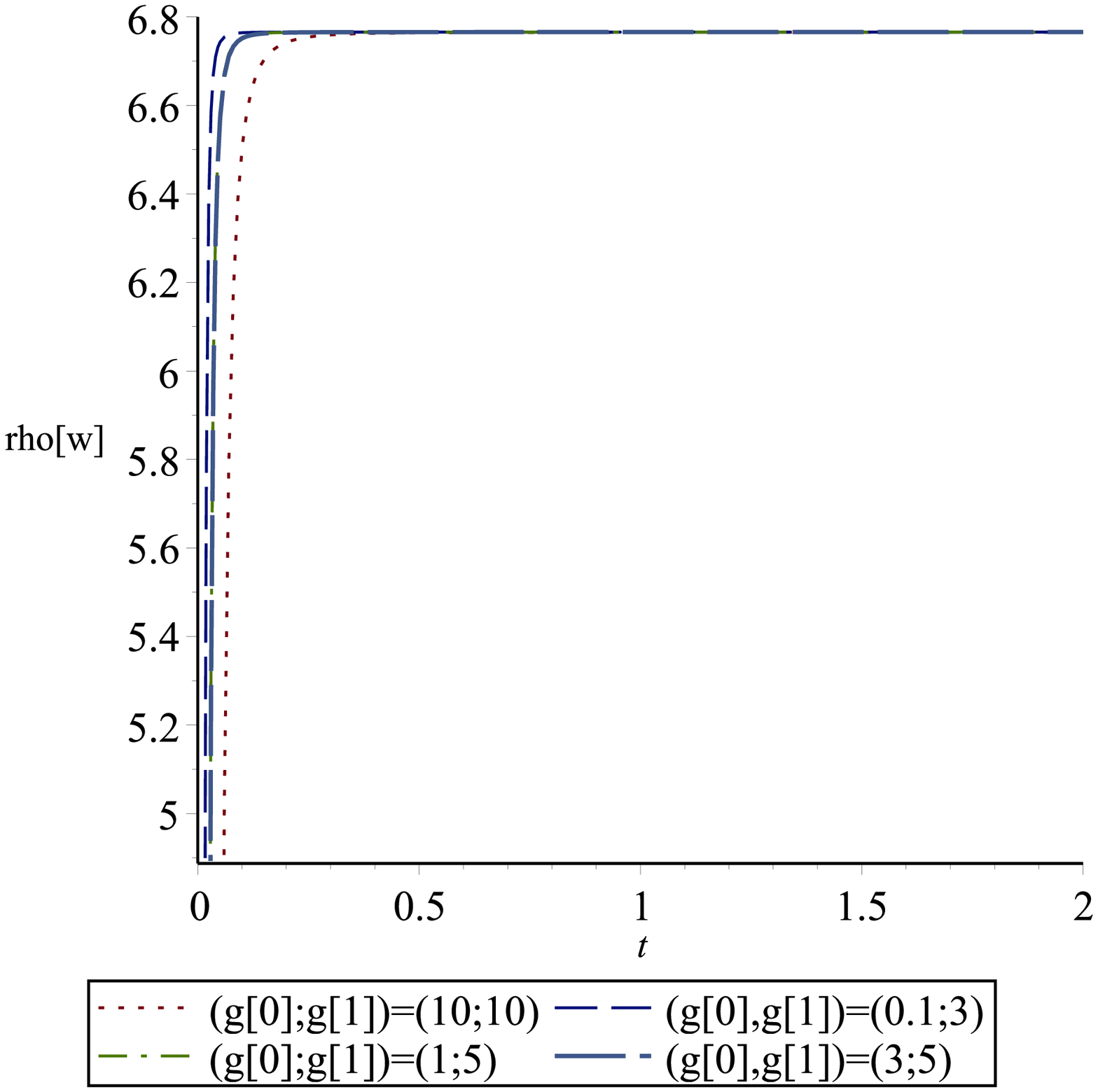} \label{e3}&
\includegraphics[height=7cm,width=7cm]{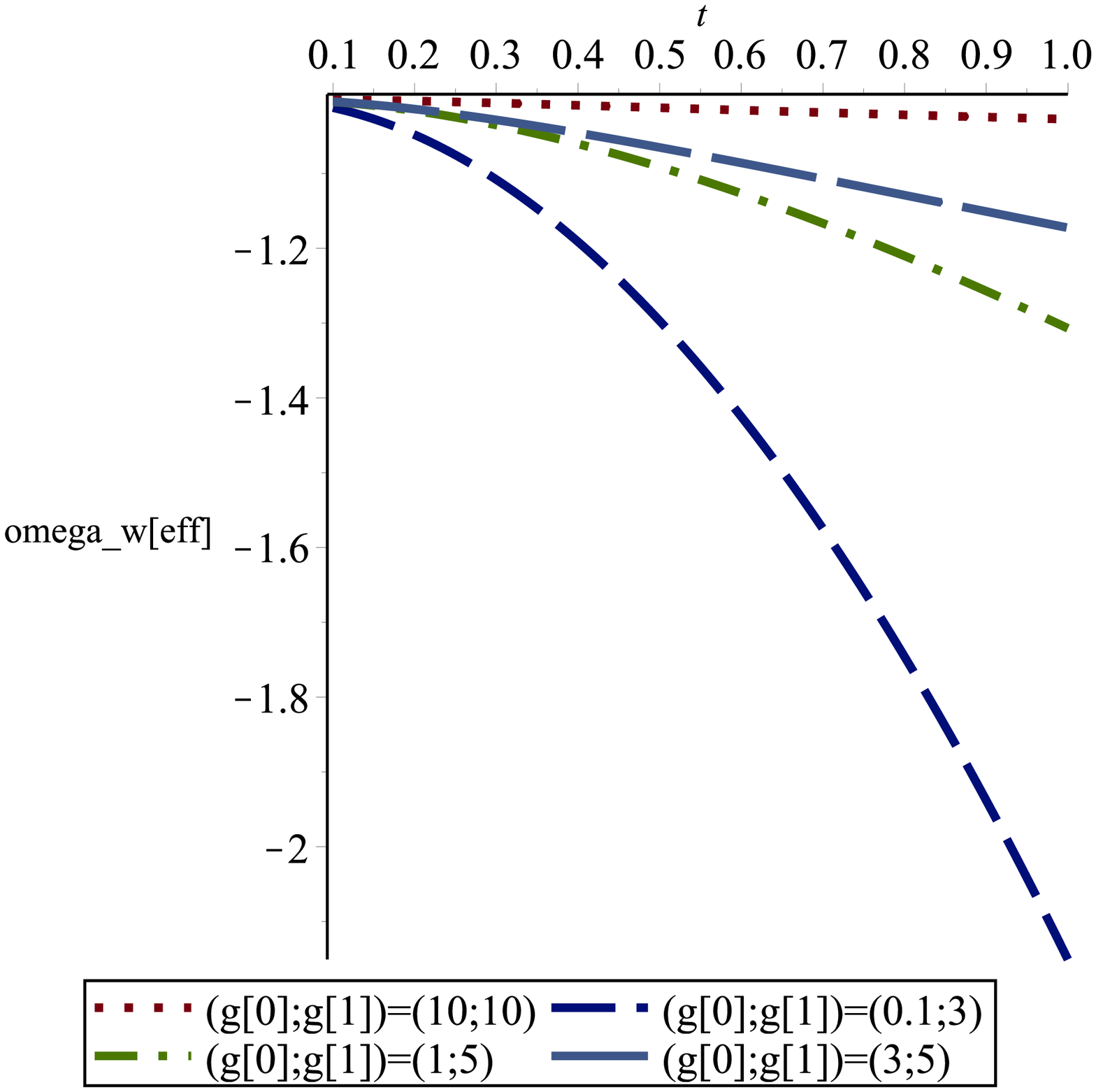} \label{e4}
\end{tabular}
\caption{The graphs present the evolution of $\rho_{w}$ (left panel) and $\omega_{eff}$ (right panel) versus $t$ in an universe composed of dark energy and Van der Waal fluid, interacting, for $\omega_{w}=0.5,E=10^{-10}$.}
\label{fig3}
\end{figure}
\newpage
\section{Interaction between dark energy and Chaplygin gas}\label{sec6}
The Chaplygin gas is characterized by a pressure that can be defined as \cite{Gorini,Bilic}
\begin{eqnarray}
 p_{Ch}=-\frac{A}{\rho_{Ch}},
\end{eqnarray}
$A$ positive constant. The equations of interaction between dark energy and Chaplygin gas are given by 
\begin{eqnarray}
 \dot{\rho}_{DE}+3H(p_{DE}+\rho_{DE})=-Q,\\
 \label{contCh}
 \dot{\rho}_{Ch}+3H(p_{Ch}+\rho_{Ch})=Q,\\
 \label{deCh}
 \dot{H}=-\frac{1}{2}(\rho_{DE}+p_{DE}+p_{Ch}+\rho_{Ch}).
\end{eqnarray}
As in the previous section, technical section \ref{sec4} is used here. And it leads to the following solutions
\begin{eqnarray}
 \rho_{Ch}=-\frac{ET^{2}}{2}+C_{8},\\
 f(T)=\frac{ET^{2}}{3}-T+\frac{C_{9}\sqrt{-T}}{2}+C_{10}.
\end{eqnarray}
Assuming $\dot{H}(t_{0})=0$ and $p_{DE}(t_{0})=-\rho_{DE}(t_{0})$, from equation (\ref{dew}) one has $p_{Ch}(t_{0})=-\rho_{Ch}(t_{0})$ and
$C_{8}=-18EH_{0}^{4}+\sqrt{A+9EH_{0}^{4}\left(1-36EH_{0}^{4}\right)}$. From equation (\ref{fr1}), it comes that 
$C_{10}=2C_{8}$. Then 
\begin{eqnarray}
 \omega_{eff}=-1-\frac{1}{T}\left(C_{10}+8E\dot{H}T-ET^{2}+\frac{4A}{ET^{2}-2C_{8}}\right),\\
 \Omega_{eff}=1.
\end{eqnarray}
\begin{figure}[htbp]
\centering
\begin{tabular}{rl}
\includegraphics[height=7cm,width=7cm]{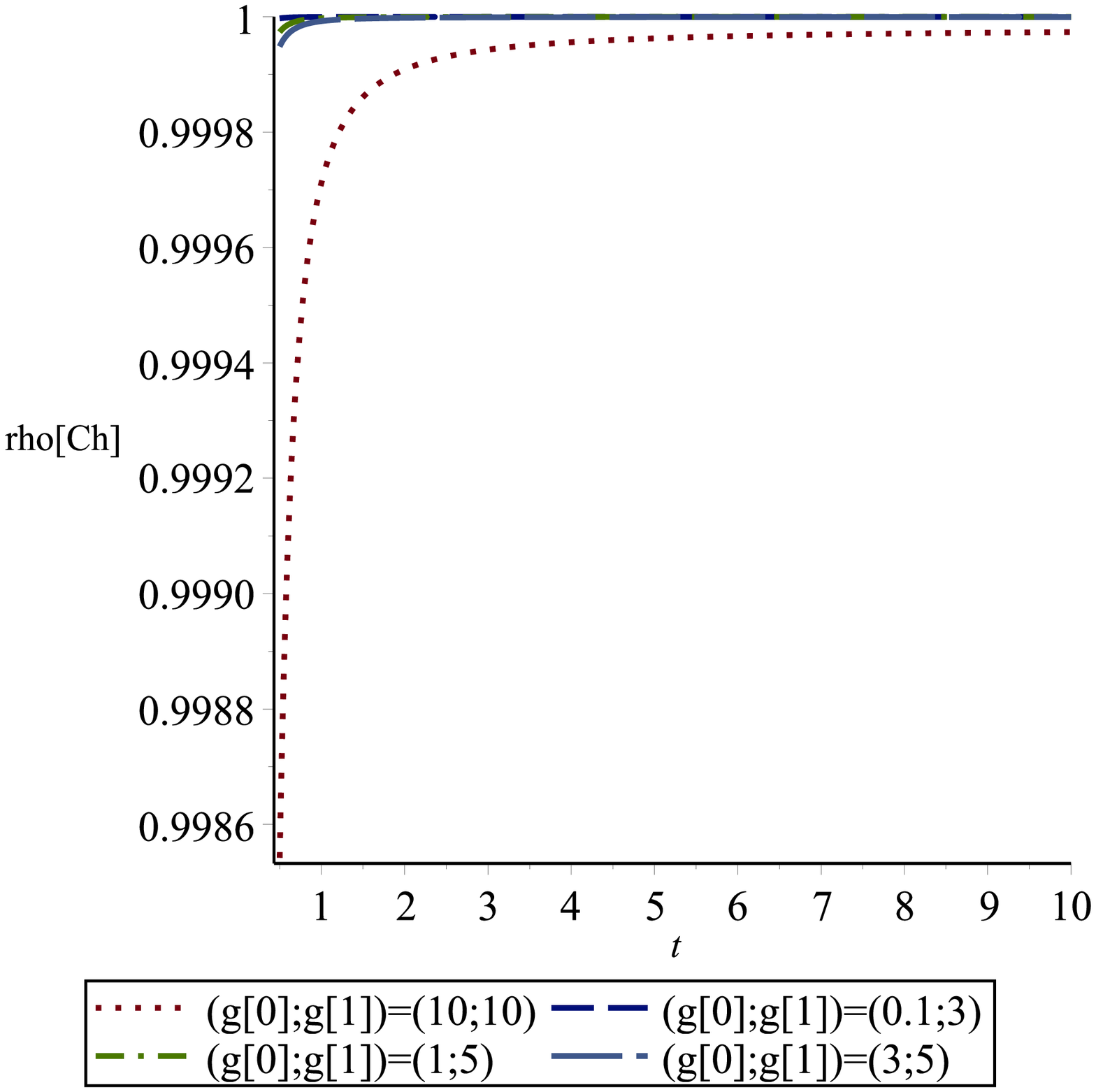} \label{e5}&
\includegraphics[height=7cm,width=7cm]{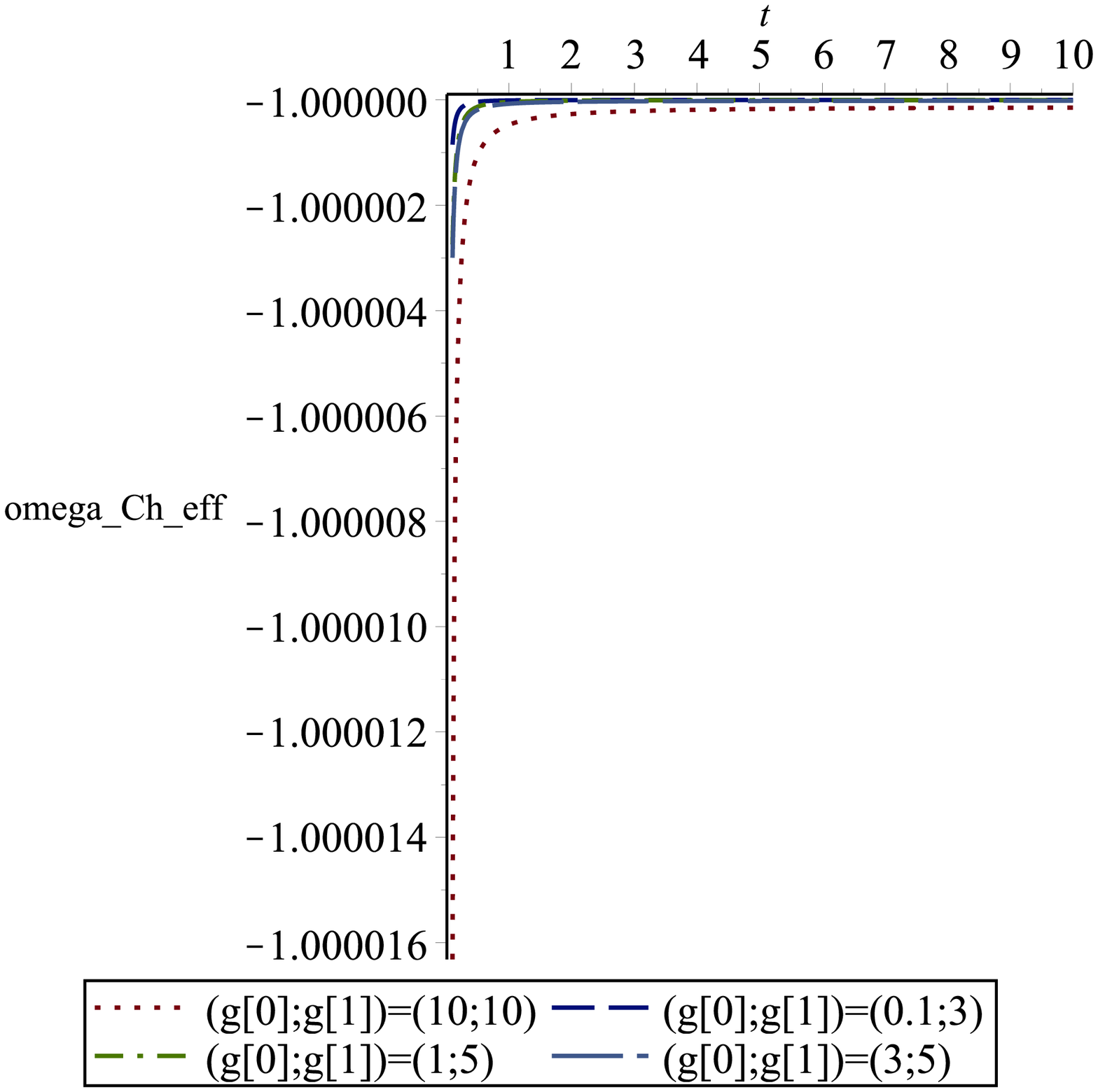} \label{e6}
\end{tabular}
\caption{The graphs present the evolution of $\rho_{Ch}$ (left panel) and $\omega_{eff}$ (right panel) versus $t$ in a universe composed of dark energy and Chaplygin gas interacting, for $A=1,E=10^{-10}$.}
\label{fig4}
\end{figure}
From Figure \ref{fig4}, we see that the energy density of Chaplygin gas $\rho_{Ch}$ tends to a constant which can be regarded as a cosmological constant because on the other side, the effective equation of state parameter $\omega_{eff}$ approaches -1.
\section{Conclusion}\label{sec7}
This paper is devoted to the study of interaction of the dark energy with some fluids in the fralework of $f(T)$ theory of gravity were $T$ denotes the torsion scalar. As fluid interacting with the dark energy, we undertake the ordinary matter fluid, the dark matter, the Van der Walls fluid and Chaplygin gas. The each fluid, an expression of the pressure is assumed, according to the previous cosmological research, and the gravitational action is found through the reconstruction scheme, within more general consideration of the expansion.
\par
In a first step the consider the interaction between the dark energy the dark matter where these mysterious components are assumed as governing the gravitational effect. Then, from the reconstruction of the Lagrangian action our result shows that the parameter of equation of state related to the energy read  $\omega_{eff}=-1$, corresponding to a universe driven by the cosmological constant $\omega_{\Lambda}=-1$. In the second step, the reconstruction of Lagrangian density $f(T)$ in a universe dominated by dark energy and ordinary matter, where dark energy is considered to be of geometric origin, is considered. The energy density of ordinary matter obtained with this approach decreases over time (Figure \ref{fig2}), in perfect agreement with the observational data.  Moreover still in the spirit of studying the interaction, the reconstruction of the action $f(T)$ in a universe composed of dark energy and van der Waals fluid  is performed leading to a universe plunging into a phantom phase ($\omega_{eff}<-1$) figure (\ref{fig3}). Finally we consider a universe dominated by dark energy and Chaplygin gas and pass to the reconstruction of the associated gravitational action; our results show that the energy density of Chaplygin gas $\rho_{Ch}$ tends to a constant while the effective equation of state parameter $\omega_{eff}$ tends to -1. This result  allows to interpret the energy density of Chaplygin gas constant of being the cosmological constant for large cosmic time.\par

\vspace{0.5cm}
{\bf Acknowledgement}: S. B. Nassur thanks DAAD for financial support. 

\end{document}